\def\ltsim{\lower3pt\hbox{$\, \buildrel < \over \sim \, $}}
\def\gtsim{\lower3pt\hbox{$\, \buildrel > \over \sim \, $}}

\def\be{\begin{equation}}
\def\ee{\end{equation}}
\def\ba{\begin{eqnarray}}
\def\ea{\end{eqnarray}}
\def\ga{\mathrel{\raise.3ex\hbox{$>$\kern-.75em\lower1ex\hbox{$\sim$}}}}
\def\la{\mathrel{\raise.3ex\hbox{$<$\kern-.75em\lower1ex\hbox{$\sim$}}}}
\newcommand{\sect}[1]{\section{#1}\setcounter{equation}{0}}

\documentstyle[aps,manuscript,epsf]{revtex}

\input epsf
\begin{document}

\begin{titlepage}

\begin{center}

\vspace{0.5cm}

\Large{\bf  Assisted Inflation in Randall-Sundrum Scenario
}\\

\vspace{10mm}

\normalsize{Yun-Song Piao$^{a,c}$, Winbin Lin$^{a}$, Xinmin Zhang$^{a}$ and Yuan-Zhong Zhang$^{b,c}$} \\

\vspace{6mm}

{\footnotesize{\it
 $^a$Institute of High Energy Physics, Chinese
     Academy of Sciences, P.O. Box 918(4), Beijing 100039, China\\
 $^b$CCAST (World Lab.), P.O. Box 8730, Beijing 100080}\\
 $^c$Institute of Theoretical Physics, Chinese Academy of Sciences,
      P.O. Box 2735, Beijing 100080, China \footnote{Mailing address, 
      Email: yspiao@itp.ac.cn}
\\}

\vspace*{5mm}
\normalsize
\smallskip
\medskip

\smallskip
\end{center}
\vskip0.6in

\centerline{\large\bf Abstract}

{We extend the Randall-Sundrum(RS) model by
adding a fundamental scalar field in the bulk,
then study the multi-field assisted inflationary solution on brane.
We will show that this model 
satisfies not only the observation, 
but also provides a solution to hierarchy problem. Furthermore in comparison
with the chaotic inflation model with a single 
field in four space-time dimension, the parameters in our model 
required for a successful inflation are natural.
}

\vspace*{2mm}

\end{titlepage}

\sect{Introduction}

Over the past two years the Randall and Sundrum (RS) \cite{RS} model has 
received much attention.
In the RS scenario two
3-branes
with opposite
tension are taken to sit at the fixed points of an $S^1/Z_2$ orbifold
with
$AdS_5$ bulk geometry. It is shown that
the gravity is localized on the brane
of positive
tension and the exponential warp factor in the space-time
metric generates a
scale hierarchy on the brane of negative tension. In this paper we will 
study the cosmological implications of the RS model. In the literature the 
related topics about the cosmological solution of the brane-world
picture \cite{low,CGRT,LPPL} have been discussed intensively
in the recent years.

In modern cosmology
inflation plays an important role. Recently the
observations by BooMERANG and MAXIMA on the location of the first
peak in the Microwave Background Radiation anisotropy strongly support for
a flat universe and favors the predictions of
inflation. In spite of these remarkable success of the inflation
models, 
there are some serious difficulties associated with the
fine-tuning of the model parameters.
To overcome these difficulties 
Liddle et al. \cite{lms} recently(see also Ref. \cite{MW})
have proposed
an assisted inflation model in which multi-scalar fields are introduced to
drive the inflation
cooperatively.

In this paper we propose a specific assisted inflation model
by introducing a fundamental bulk scalar field in the
RS scenario, and we will show that due to the large number of
the Kaluza-Klein (KK) modes of the bulk scalar field, this model in the
absence of
the fine-tuning of the model parameters satisfies the
observation, meanwhile it keeps the nice features of the RS scenario in
solving the hierarchy problem. We have noticed that the inflationary 
solutions
in the RS scenario in\cite{mw1},
however, the fine-tuning of the model
parameters is still required in these models.
The paper is organized as follow:
in section 2 we introduce the model and study the inflationary
solution. Section 3 is our conclusion.

\sect{The Model}

Our model is an extension of the
RS model \cite{RS}
by adding a fundamental scalar field
$\tilde \Phi$ with mass $m$ in
the bulk as shown in $S_{bulkscalar}$ below. The action of the model is 
$S= S_{bulk} + S_{brane} + S_{bulkscalar}$ where 
\be
S_{bulk}=\int d^4x dy
\sqrt{\tilde{g}}\left({1\over 2\tilde{\kappa}^2} \tilde{R}-\tilde{\Lambda}\right),
\ee
\be
S_{brane}=\int d^4x \sqrt{-g}({\cal L}_{m,0}-V_0)|_{y=0}
+\int d^4x \sqrt{-g}({\cal L}_{m,l}-V_l)|_{y=l},
\ee
\be
S_{bulkscalar}=\int d^4 x dy {\sqrt{\tilde{g}}}
\left({1\over 2}\partial_A\tilde{\Phi}\partial^A\tilde{\Phi} -
{1\over 2}m^2\tilde{\Phi}^2\right) .
\ee
The RS model described by action $S = S_{bulk} + S_{brane}$ in Eqs. (1) 
and (2)
consists of two 3-branes in an $AdS_5$
space with a negative cosmological constant $\Lambda$,
 and two 3-branes are assigned to have opposite brane tensions
and located at the
orbifold fixed points $y=0$ and $y=l$.
Note that
the variables with ``$\, \tilde {}$ " in Eqs.(1)-(3) 
represent the five dimensional one and $x^A=(x^\mu,y)$.
The distance between the $3$-branes is assumed
to be stabilized by the mechanisms 
in \cite{GW1,PZZ}.

In general the
five dimensional metric that respects the 
cosmological principle on the brane can be written as 
\be
ds^{2}=-n^{2}(t,y) d t^{2}+a^{2}(t,y)d{\mathbf{x}^2}+b^{2}
(t,y)dy^{2}.
\ee
Following Ref.\cite{CGRT}, we
consider the energy density of the scalar field in the bulk and the matter in
both $3$-branes much less than the bulk cosmological constant and 
the brane tension, so that this metric can be linearized around the RS 
solution \cite{CGRT} \begin{eqnarray}
a(y,t)&=&a(t) \exp{(-\sigma(y)b(t))} \left(1+ 
\delta a(y,t) \right) \nonumber \\ 
n(y,t)&=&\exp{(-\sigma(y)b(t))}\left(1+\delta n(y,t) \right) 
\nonumber \\
b(y,t)&=&b(t) \left(1+ \delta b(y,t)\right) .
\end{eqnarray}
where $\sigma(y)=ky$ with $k^2=(-\Lambda \tilde{\kappa}^2)/6$ the bulk 
curvature, and 
$\delta n, \delta a, \delta b\sim O(\rho_0,\rho_l)$ with $\rho_0, \rho_l$ 
the energy density of matter on the two branes since in the limit of 
$\rho_0,\rho_l\rightarrow 0$ the RS solution should be recovered. 
Therefore, 
without loss of generality, $b=1$ is taken in this paper and the metric
in Eq.(4) is reduced to 
\be
ds^2=e^{-2\sigma(y)}g_{\mu\nu}dx^{\mu}dx^{\nu} - dy^2,
\ee
where $g_{\mu\nu}=dt^2-a^2(t)d{\mathbf x}^2$ is the
flat Friedmann-Robertson-Walker (FRW) metric. And one can see that 
this metric gives rise to
a expansion rate observed by the observer of two $3$-branes 
identically \cite{CGRT,LPPL}. 

Assisted inflation model requires multi-scalar fields \cite{lms,MW}. 
In our model,
upon compactification the scalar field $\tilde \Phi$ gives rise to a set of
effective scalars in the four dimension and the KK modes of the bulk
scalar serve as inflaton in the assisted inflation scenario.
Now we closely follow the procedure in
Ref. \cite{GW}and 
work out the effective potential of the KK scalars. 

Starting with the action in Eq.(3), the metric in Eq.(6) and making use of
the integration by parts we can rewrite
the action of the bulk scalar field  
as
\be
S_{bulkscalar}={1\over 2}\int d^4 x\int dy \left(e^{-2\sigma(y)}
\partial_\mu{\tilde{\Phi}} \partial^\mu {\tilde{\Phi}} + 
{\tilde{\Phi}}{\partial\over \partial y}
(e^{-4\sigma(y)}{\partial {\tilde{\Phi}}\over\partial y})
-m^2 e^{-4\sigma(y)}{\tilde{\Phi}}^2\right).
\ee
Defining 
$\Psi_i(y)$ a set of scalar fields in the fifth dimension which satisfies
$\int dy \Psi_i(y) \Psi_j(y) \exp{(-2\sigma)}=\delta_{ij}$ and
\be
{\partial\over \partial y} 
\left(e^{-4\sigma(y)}{\partial \Psi_i\over \partial y} \right) - 
m^2 e^{-4 \sigma(y)} \Psi_i = -m_i^2 e^{-2\sigma(y)} \Psi_i.
\ee
We expand  
$\tilde{\Phi}(x,y)$ as a sum over modes, 
{\it i.e.} $\tilde{\Phi}(x,y)=\sum_i \Phi_i(x) \Psi_i(y)$. Then, 
after integrating the five dimensional action over the extra dimension $y$, 
we obtain
an effective four dimensional action which describes the evolution of
the scale factor $a(t)$ on the two $3$-branes 
\be
S=\int d^4x \sqrt{-g}{1\over 2\kappa^2} R+{1\over 2}\sum_i\int d^4x\sqrt{-g}
(\partial_{\mu}\Phi
_i\partial^{\mu}\Phi_i-m^2_i\Phi_i^2).
\ee
In Eq.(9) ${1\over 2\kappa^2}={1\over 2k\tilde{\kappa}^2}(1-e^{-2kl})$, 
from which one can see that for $e^{kl}\gg 1$, $M\sim k\sim 
m_p$ with $m_p$ the four dimensional Planck scale.

The solution of Eq. (8) is 
\be
\Psi_i(y)=a_{i\nu} e^{2\sigma(y)}\left(J_\nu \left({m_i\over k} e^{\sigma(y)}\right)
+ b_{i\nu} N_\nu \left({m_n\over k} e^{\sigma(y)}\right)\right),
\ee
where $a_{i\nu}$ is a normalization factor, $J_\nu $ and $N_\nu$ are the Bessel 
function. The jump conditions for two 3-branes 
give rise to two relations which can be used to solve for $m_i$ and 
$b_{i\nu}$. 
As usual,
the bulk field $\tilde{\Phi}(x,y)$ upon compactifications manifests 
itself to a observer in the four space-time dimension
as an infinite ``tower'' of scalars with masses $m_i$. And their mass 
spectrum satisfy equation \be
2 J_\nu (\frac{m_i e^{kl}}{k}) + \frac{m_i e^{kl}}{k}
 J'_\nu (\frac{m_i e^{kl}}{k})=0,
\ee
with $\nu=\sqrt{4+{m^2\over k^2}}$.

  From Eq. (11), we will have a relation
between $m_i$ and
$m$. In Fig. 1 we plot the KK scalar mass spectrum as function of the 
bulk scalar mass m. One can see that for $m\leq 0.1k\sim 0.1m_p$, the 
dependence of
the mass spectrum $m_i$ on the mass scale $m$ of the bulk scalar field
is very weak.
Therefore, we approximate the mass formula as
\be
m_i\simeq \xi_1(1+(i-1)\pi)k e^{-kl}.
\ee

In the usual single-field inflation in four
space-time dimension
or the assisted inflation
in the large extra dimension \cite{kl},
the scalar mass is determined by the COBE observation,
so one has a strong
constrain on the mass
 $m \sim 10^{-6}m_p$. We will show
 below that this will not be the case in our model.

For the infinite "tower" of the scalar the 
natural cut off of the mass spectrum is the Plank scale $m_p$. With such 
a cutoff  we will have the total number of the KK fields
\be
{\cal{N}}_{tot}\approx {\exp{(kl)}\over \pi\xi_1}.
\ee

Now we study the inflationary solution of this
scenario. Given the action (9) we have the
equations of motion
\be
H^2={4\pi\over 3m_p^2}\sum_i(\dot{\Phi}_i^2+m_i^2\Phi_i^2),
\ee
\be
\ddot{\Phi}_i+3H\dot{\Phi}_i+m_i^2\Phi_i=0.
\ee
In the slow-rolling approximation, one drops the second derivative term 
in Eq. (15), and obtains that
\be
{\Phi_i(t)\over \Phi_i(0)}=\left({\Phi_1(t)\over \Phi_1(0)}\right)^{m_i^2/m_1^2},
\ee
where the Hubble parameter $H\equiv\dot{a}/a$ represents
the expansion rate of two
$3$-branes and $\Phi_i(0)$ is the initial values of $\Phi_i(t)$.

To calculate the amplitude of adiabatic perturbation generated
during inflation and the corresponding index of spectrum, we use the usual
formula for the perturbations of the multi-field in the four dimensional
space-time.
Following Refs. \cite{SS} and \cite{LR}, we have for our model
the e-folding number 
\be
N={2\pi \over m_p^2}\sum_i \Phi_i^2,
\ee
and the amplitude of adiabatic perturbation 
\be
\delta_{H}={H\over 2\pi} 
\sqrt{\sum_i\left({\partial N \over \partial \Phi_i}\right)^2}.
\ee

In general the  initial values of the multi scalars in the assisted 
inflation model could be taken to be different, however for simplicity as 
what has been done in
\cite{kl} we assume the same for the initial values of 
all KK
fields. With such an assumption, the discussions will be 
much simple. After substituting (16) into (17) and (18), we obtain that
 \be
N={2\pi\over m_p^2}\Phi_1^2(0)\sum_i r_i,
\ee
\be
\delta_{H}=\sqrt{{4\pi\over 3}}{m_1\Phi_1^2(0)\over m_p^3}
\sqrt{(\sum_i \mu_i r_i)(\sum_i r_i)},
\ee
where $\mu_i\equiv (m_i/m_1)^2$ and
$r_i \equiv (\Phi_1(t)/\Phi_1(0))^{2\mu_i}$. 

During the inflation the KK fields with mass less than the Hubble parameter
drive the inflation, however, those with mass larger are diluted
quickly. Quantitatively, for
$m_{\cal{N}}^2\leq {H}^2 < m_{{\cal{N}}+1}^2$, we have
\be
\mu_{\cal{N}}\leq {4\pi\Phi_1^2(0)\over 3m_p^2}\sum_{i=1}^{{\cal{N}}}
\mu_i r_i <
\mu_{{\cal{N}}+1},
\ee
where ${\cal{N}}$ denotes the number of KK fields which are
in the process of
slow-rolling
during the inflation.
Because that
the difference between $m_{\cal{N}}$ and
$m_{{\cal{N}}+1}$ is the same order as $m_{\cal{N}}$ (see Eq. (12)), {\it 
i.e.} $\mu_{\cal{N}}\sim\mu_{{\cal{N}}+1}$,
Eq. (21) can be reduced to
\be
\mu_{\cal{N}}\simeq {4\pi\Phi_1^2(0)\over 3m_p^2}\sum_{i=1}^{{\cal{N}}}
\mu_i r_i.
\ee

Making use of the definitions of $\mu_i$ and $r_i$,
Eq. (19) and Eq. (22) can be rewritten as
\be
N={2\pi\over m_p^2}\Phi_1^2(0)\sum_{i=1}^{\cal{N}}
\left[{\Phi_1(t)\over \Phi_1(0)}\right]
^{2[1+(i-1)\pi]},
\ee
\be
[1+({\cal{N}}-1)\pi]^2\simeq {4\pi\Phi_1^2(0)\over 3m_p^2}\sum_{i=1}^{\cal{N}}
[1+(i-1)\pi]^2\left[{\Phi_1(t)\over \Phi_1(0)}\right]
^{2[1+(i-1)\pi]}.
\ee

Taking $N\simeq 60$ as required when the COBE scale exits the Hubble radius,
and eliminating
$\Phi_1(t)$ in (23) and (24), we obtain a
relation between ${\cal{N}}$ and $\Phi_1(0)$,
which is shown
in Fig. 2. One can see that the number of the KK fields participating in the
slow-rolling process increases as $\Phi_1(0)$ decreases.

Combining (19), (20) and (22), we have
\begin{eqnarray}
\delta_{H}&\simeq &{1\over \sqrt{2\pi}}{m_1\over m_p}
\sqrt{N \mu_{\cal{N}}}\nonumber\\
&\simeq &{1\over \sqrt{2\pi}}{m_1\over m_p}(1+({\cal{N}}-1)\pi)\sqrt{N}.
\end{eqnarray}
Making use of the COBE observation
$\delta_{H}\simeq 10^{-5}$,  and
$m_1\simeq\xi_1 k e^{-kl}$, we have
\be
\exp{(-kl)}\simeq {10^{-6}\over \cal{N}}.
\ee
Note that $e^{kl}\sim 10^{16}$ which is required to solve the hierarchy
problem between the Planck scale
and the electroweak scale, Eq. (26) shows the number of the KK field 
which are in slow rolling is large: ${\cal{N}}\sim 
10^{10}$. From Fig. 2, we see that in this case $ \Phi_{1}(0)\ll m_p $.

The corresponding index of spectrum is
\begin{eqnarray}
n-1&=&{d\ln{\delta_{H}^2}\over d\ln{k}}
=-{d\ln{\delta_{H}^2}\over d N}\nonumber\\
&\simeq&-{1\over N}-{2\over {\cal{N}}}{d{\cal{N}}\over d N},
\end{eqnarray}
which is close to zero {\it i.e.} a near scale-invariant spectrum.

The effective potential of the total KK fields in our model
on the 3-brane can be written as:
\be
V\simeq {1\over 2}m_1^2\Phi_1^2(0)(\sum_i^{\cal{N}} \mu_i r_i).
\ee
Substituting (22) into (28), and taking into account that 
$\mu_{\cal{N}}\equiv ({m_{\cal{N}}\over m_1})^2
\sim {\cal{N}}^2$ (notice that ${\cal{N}}\sim 10^{10} \gg 1$), we
obtain
\be
V\sim {3\over 8\pi}m_1^2 m_p^2{\cal{N}}^2 .
\ee
Since $m_1\simeq\xi_1 k e^{-kl}$ and $e^{-kl}\sim 10^{-16}$, we have
$V\sim 10^{-12} m_p^4$. This shows that  although the number of
KK fields are quite large  $ \sim 10^{10}$
 the potential energy $V$ is still far below the Planck
scale.
Therefore, the assumption that energy density
of matter on the 
$3$-brane is much less than both the bulk
cosmological constant and the brane tension is reasonable and
the quadratic term of energy density is not important in our consideration 
\cite{low}.

\sect{Conclusion}

In summary, we have considered a model by including a bulk scalar
field in the RS scenario and studied the inflationary solution. We
have calculated the amplitude of scalar perturbation generated
during inflation and the index of the spectrum.
Our results show that the initial value of the inflaton is not
required to be higher than the Planck scale and the scalar boson mass
$m$ is not required to be much less than the Plank scale, furthermore
this model solves the
hierarchy problem. All of these features are attractive compared with the chaotic
inflation model of single scalar field
in the four dimensional space-time in which the initial value of the
scalar field must be higher than the Planck scale to make inflation
happen and its mass must be far less than the Planck scale to satisfy
the COBE observation.

\textbf{Acknowledgments}

We thank R. Brandenberger
for stimulating discussion.
This work is supported in part by National Natural Science Foundation of
China under Grant Nos. 10047004 and 19835040, and also supported by
Ministry of Science and Technology of China under Grant No NKBRSF G19990754.

\newpage

\begin{figure}
\epsfxsize=3.0 in \epsfbox{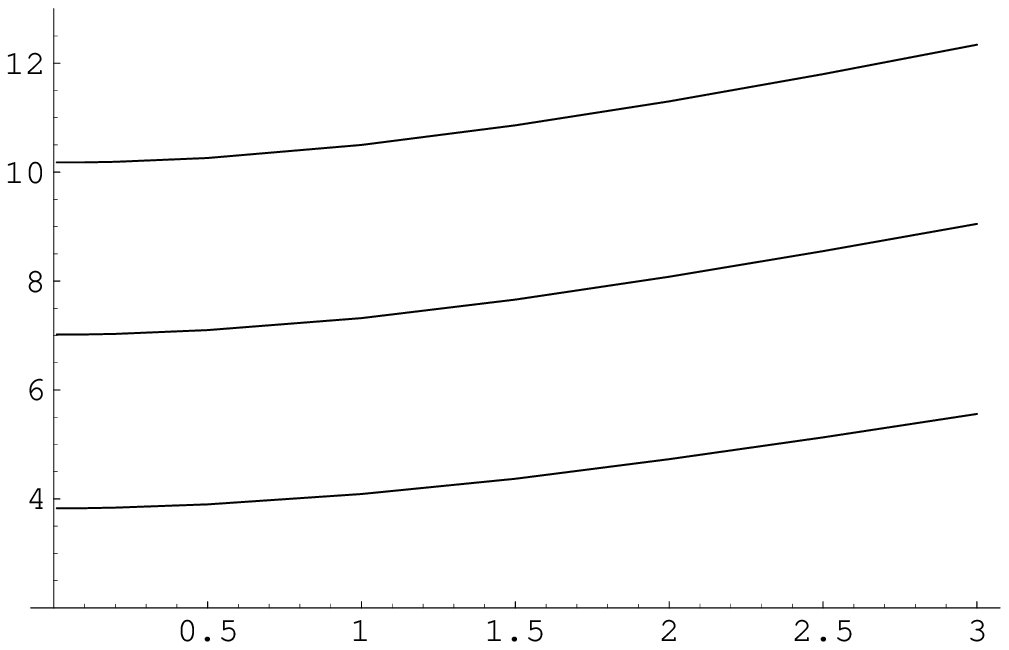}
\caption { Plot of KK field masses vs the bulk scalar mass.
The $x$-axis is ${m\over k}$,
and the $y$-axis is ${m_i\over k}e^{kl}$. The three curves
corresponds to the three lightest KK modes.
}
\end{figure}

\vskip 1cm

\begin{figure}
\epsfxsize=3.0 in \epsfbox{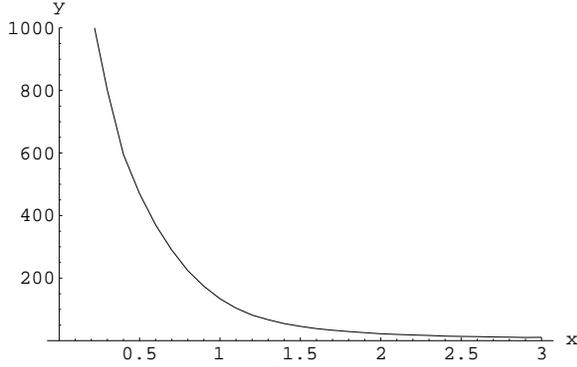}
\caption {Plot of $\cal{N}$ as function of $\Phi_1(0)$.
The $x$-axis is $\Phi_1(0)$ in
unit of $m_p$, the $y$-axis is $\cal{N}$. In the numerical
calculation we take
 $N=60$.
}
\end{figure}

\end{document}